\documentclass[preprint,showpacs,preprintnumbers,amsmath,amssymb,nofootinbib,showkeys,aps]{revtex4}
\usepackage{graphicx}
\usepackage{dcolumn}
\usepackage{bm}
\usepackage{latexsym}
\usepackage{epsfig}
\usepackage{bm}

\newcommand{\be}{\begin{equation}}
\newcommand{\ee}{\end{equation}}
\newcommand{\beqq}{\setlength\arraycolsep{2pt}\begin{eqnarray}}
\newcommand{\eeqq}{\vspace{0cm} \end{eqnarray}}
\newcommand{\bea}{\begin{eqnarray}}
\newcommand{\eea}{\end{eqnarray}}

\begin{document}

\title{CCDM model from quantum particle creation: constraints on dark matter mass}

\author{J. F. Jesus} \email{jfjesus@itapeva.unesp.br}
\affiliation{Universidade Estadual Paulista ``J\'ulio de Mesquita Filho'' -- Campus Itapeva \\
Rua Geraldo Alckmin 519, 18409-010, Vila N. Sra. de F\'atima, Itapeva, SP, Brazil}

\author{S. H. Pereira} \email{shpereira@gmail.com}
\affiliation{Universidade Estadual Paulista ``J\'ulio de Mesquita Filho'' \\
 Departamento de F\'isica e Qu\'imica -- Faculdade de Engenharia de Guaratinguet\'a \\ Av. Ariberto Pereira da Cunha 333, 12516-410, Pedregulho, Guaratinguet\'a, SP, Brazil}
   

\keywords{dark matter theory, particle physics - cosmology connection }

\begin{abstract}
In this work the results from the quantum process of matter creation have been used in order to constrain the mass of the dark matter particles in an accelerated Cold Dark Matter model (Creation Cold Dark Matter, CCDM). In order to take into account a back reaction effect due to the particle creation phenomenon, it has been assumed a small deviation $\varepsilon$ for the scale factor in the matter dominated era of the form $t^{\frac{2}{3}+\varepsilon}$. Based on recent $H(z)$ data, the best fit values for the mass of dark matter created particles and the $\varepsilon$ parameter have been found as $m=1.6\times10^3$ GeV, restricted to a 68.3\% c.l. interval of ($1.5<m<6.3\times10^7$) GeV and $\varepsilon = -0.250^{+0.15}_{-0.096}$ at 68.3\% c.l. For these best fit values the model correctly recovers a transition from decelerated to accelerated expansion and admits a positive creation rate near the present era. Contrary to recent works in CCDM models where the creation rate was phenomenologically derived, here we have used a quantum mechanical result for the creation rate of real massive scalar particles, given a self consistent justification for the physical process. This method also indicates a possible solution to the so called ``dark degeneracy'', where one can not distinguish if it is the quantum vacuum contribution or quantum particle creation which accelerates the Universe expansion.

\end{abstract}

\maketitle

\section{Introduction}

The idea of an accelerating universe is indicated by type Ia Supernovae observations \cite{SN,ast05}, and from a theoretical point of view, a hypothetical exotic component with large negative pressure may drive the evolution in an accelerating manner. This exotic component is usually termed quintessence or dark energy and represents about 70\% of the material content of the universe (see \cite{reviewDE} for a review). The simplest example of dark energy is a cosmological constant $\Lambda$ \cite{weinberg,padmana,martin}, and recently the 7 year WMAP \cite{wmap7} have indicated no deviation from the standard $\Lambda$-Cold Dark Matter ($\Lambda$CDM) model.

Nevertheless, recently a new kind of accelerating cosmology  with no dark energy has been investigated in the framework of general relativity, called Creation Cold Dark Matter (CCDM) \cite{lima01,limafernando,debnath,RoanyPacheco10,LJONote,JesusEtAl11}. In this scenario the present accelerating stage of the universe is powered by the negative pressure describing the gravitational particle creation of cold dark matter particles, with no need of a dark energy fluid. Such phenomenological models are constructed by giving a specific form to the creation rate but with no further physical grounds. Here we present a creation rate derived from a quantum mechanical particle creation process for real massive scalar fields and we constrain the mass of the field in order to satisfy the observational data. This way, our model has the advantage to be self consistent from first principles. As far as we know it is the first time that a real quantum process of matter creation is analysed in the context of CCDM model. Our analysis can be seen as a first approximation of the full quantum creation process, as it takes into account the effect of creation into the evolution of the scale factor as a second order effect, which can be understood as a kind of back reaction effect in the expansion.

The nature and origin of the dark matter (DM) is still a mystery (see \cite{DMrev,bookDM} for a review and \cite{tenpoint} for a ten-point test that a new particle has to pass in order to be considered a viable dark matter candidate), thus such accelerating cold dark matter models have the advantage of explaining the present stage of acceleration of the Universe and the origin of the dark matter, two of the main challenges of the present cosmology. From a quantitative point of view, the search for candidates to particles of DM has increased in the context of new theories beyond the standard model of particle physics\footnote{For a good reference, see chapters 7 to 12 of \cite{bookDM}. According to the theory of Big Bang Nucleosynthesis
(BBN), the simplest and most plausible form of non-baryonic cold dark
matter particle are the WIMPs (Weakly Interacting Massive Particles).
The determination of the relic density for the WIMPs depends on the
evolution of the universe before BBN, and if its mass is $m>100$ MeV
it could freeze out before BBN and would be the earliest remnants of
cold dark matter in the universe. A specific kind of WIMP particle is
the WIMPZILLA, a very massive relic from Big Bang, that might be
produced at the end of inflation by the gravitational creation of
matter during the accelerated expansion. If its mass is about
$10^{13}$ GeV it might be the dark matter in the universe.
Supersymmetric particles are also candidates to dark matter in models
beyond the Standard Model, since that cold dark matter are predicted
in a very natural way in models of supersymmetry. The Lightest
Supersymmetric Particles (LSP) as neutralino and gravitino are some
kinds of particles expected from these models. For LSP particles a
typical mass range is 50-1000 GeV. Theories based on Kaluza-Klein (KK)
parity in Universal Extra Dimension (UED) model leads to masses of
about 500-1500 GeV. For theories based on warped extra dimensions, the
Lightest $Z_3$ Particle (LZP) model predicts a mass of about 20 GeV -
1 TeV. There are also non-WIMP dark matter candidates, as the axions
and sterile neutrinos. Axions have a double motivation in
astroparticle models since they solve the strong CP problem and are
also candidate to dark matter. Although being a good candidate,
several constraints imposed by different experiments and models ties
the axion mass to about 1 eV or $10^{-3}$ eV, a very tiny mass. The
sterile neutrino is also very light, with mass of about 1-100 keV. From these models we see that the mass of dark matter particles can be accommodated in a large spectrum of values for different theories.}

The first 
self-consistent macroscopic formulation of the matter creation process 
was presented in \cite{Prigogine} and formulated in a covariant form in \cite{CLW}. In comparison to the standard equilibrium 
equations, the process of creation  at the expense of the gravitational field is described by two new ingredients: a balance 
equation for the particle  number density and a negative  
pressure term in the stress tensor. Such quantities 
are related to each  other in a very definite way by the second law of 
thermodynamics. In  particular, the creation pressure depends on the creation rate and may 
operate, at level of Einstein's equations, to prevent either a space-time 
singularity \cite{Prigogine,AGL} or to generate an early inflationary 
phase \cite{LGA}. The quantum process of particle creation has been studied by several authors \cite{davies,fulling,grib,mukh2,partcreation,staro,pavlov01,pavlov02,gribmama02,fabris01} in the last five decades, and the results are well known. Recently the gravitational fermion production in inflationary cosmology was revisited for both the large and the small mass regimes \cite{chung}.

In this article we analyse results for the rate of creation of real massive scalar particles needed to accelerate the universe as currently observed, assuming that the particles created are dark matter particles. Based on recent observational data, we constrain the value of the mass of the dark matter particles. 

The paper is organised as follows. In Section II it is presented the macroscopic effects of particle creation in CCDM model. In Section III it is presented only the results for real massive scalar particle creation in Friedmann models based on \cite{staro}. The general theory of particle creation is briefly presented in the Appendix for sake of generality. The main results are in Section IV where it is constrained the dark matter mass using the results of the previous section into the CCDM model equations of Section II. In Section V it is compared the quantum CCDM model with some phenomenological models and $\Lambda$CDM. The conclusions are presented in  Section VI.

\section{Macroscopic effects of particle creation}

As we said before, the macroscopic effect of particle creation in an expanding universe brings the possibility of an accelerated expansion, depending on the rate of creation of these particles. Let us put this on quantitative grounds.

In a homogeneous and isotropic Friedmann-Robertson-Walker background, the non-trivial Einstein's field equations for a mixture of radiation, baryons and dark matter endowed with dark 
matter creation and the energy conservation laws for each component can be written as \cite{Prigogine,CLW}
\begin{equation}
    8\pi G (\rho_{rad}+ \rho_{bar}+\rho_{dm})= 3 \frac{\dot{a}^2}{a^2} + 3 \frac{\kappa}{a^2},
		\label{fried}
\end{equation}

\begin{equation}
   8\pi G (p_{rad}+p_{c}) = -2 \frac{\ddot{a}}{a} - \frac{\dot{a}^2}{a^2}- \frac{\kappa}{a^2},
\end{equation}

\begin{equation}\label{rhorad}
      \frac{\dot{\rho}_{rad}}{\rho_{rad}} + 4 \frac{\dot{a}}{a} = 0,
\end{equation}

\begin{equation}\label{rhobar}
      \frac{\dot{\rho}_{bar}}{\rho_{bar}} + 3 \frac{\dot{a}}{a} = 0,
\end{equation}

\begin{equation}\label{rhodm}
      \frac{\dot{\rho}_{dm}}{\rho_{dm}} + 3 \frac{\dot{a}}{a} = \Gamma,
\end{equation}
where an over-dot means time derivative, $\rho_{rad}$, $\rho_{bar}$ and $\rho_{dm}$ are  the radiation, baryonic and dark matter energy densities, $p_{rad}$ and $p_{c}$ are the radiation and creation pressure and $\Gamma$ is the dark matter creation rate of the process. The creation pressure 
$p_{c}$ is defined in terms of the creation rate and other physical 
quantities.  In the case of adiabatic 
matter creation, it is given by \cite{Prigogine,CLW,AGL,LGA}
 
\begin{equation}\label{CP}
    p_{c} = -\frac{\rho_{dm}}{3H} \Gamma,
\end{equation}
where $H = {\dot{a}}/a$ is the Hubble parameter.

The above expressions show how the matter creation rate,  $\Gamma$, modifies the 
evolution of the scale factor as compared to the case with no creation. Conversely,  
the cosmological dynamics with irreversible matter creation will be defined once the matter creation rate is given. 
As should be expected, by taking  $\Gamma=0$ it reduces to the FRW differential equation governing the evolution of a perfect simple fluid \cite{weib,kolb,mukh}.

\section{Particle creation in Friedmann models}

After the works of Parker \cite{parker}, the phenomenon of particle creation in cosmological context has been studied by several authors \cite{davies,fulling,grib,mukh2,partcreation,staro,pavlov01,pavlov02,gribmama02,fabris01}. One of the most interesting results from Parker's work is that exactly no massless particle is created in a radiation dominated universe, either of zero or non-zero spin. And also super-massive particles can not be created in a matter dominated universe. Among the theories of particle production by the gravitational field, two different methods are commonly used. The standard method adopted by Parker and others \cite{davies,fulling,parker} is the method of the adiabatic vacuum state, where the vacuum state is defined as one for which the lowest-energy state goes to zero smoothly as $t\to 0$ in the past and also in the future, represented by ``in'' and ``out'' states with a definite number of particles. However, this is not the case in cosmology, since the expansion of the universe can not be taken as a smooth expansion, at least in the past. This leads to the problem of particle interpretation if there are no static ``in'' or ``out'' regions. In order to solve this problem, they introduced a method of selecting those modes solutions of the field equation that come in some sense closest to the Minkowski space limit. It corresponds physically to a definition of particles for which there is minimal particle production by the changing geometry \cite{davies}. An alternative formulation of particle creation by a gravitational field that has been widely used is that of instantaneous Hamiltonian diagonalization \cite{grib, pavlov01} suggested by Grib and Mamayev \cite{gribmama02}, where the vacuum states are defined as those which minimises the energy at a particular instant of time and the Hamiltonian is constructed via the metrical energy-momentum tensor instead of the canonical one, since in general relativity the energy is obtained by means of variation of the action integral with respect to $g_{\mu\nu}$. This method also has the advantage of predicting much more created particles than the previous method. Of particular interest in this method are the works relating the gravitational particle production at the end of the inflation as a possible mechanism to produce super-heavy particles of dark matter \cite{pavlov02,superDM}. Here we restrict ourselves to this second method. The details are in the books \cite{mukh2,grib} and also in the articles \cite{partcreation,staro,gribmama02}. The main equations are derived in the Appendix.

Let us present here only the results of the above considerations to Friedmann models of the universe, where the scale factor is given by the power law
\be
a(t)=a_0 t^q\,,\hspace{1cm} a(\eta)=a_0^{1\over 1-q}(1-q)^{q\over 1-q}\eta^{q\over 1-q}\,,
\ee
where $a_0$ is a constant and represent the present scale factor of the universe, $\eta$ is conformal time, $q=1/2$ stands for radiation and $q=2/3$ for matter-type background. We also suppose that $t=0$ corresponds to $\eta=0$. We will work with units where $c=\hbar=1$. We will neglect the baryon and radiation contribution, as we are interested on the late stages of Universe expansion. We also assume that the Universe is spatially flat, except where mentioned.

The detailed discussion for particle creation of real and complex scalar field, spinor field and vector field is given in \cite{grib} for open, closed and flat models. Here we will present only the main results concerning the particle production for massive real scalar fields with conformal coupling $\xi=1/6$ in a flat space-time ($\kappa=0$). For the details of such calculations see \cite{staro}.

For real models of the universe, the influence of the spatial curvature on the expansion at $t \sim m^{-1}$ is still very small. Analytic calculations can be done in two important cases: $t<< m^{-1}$ and  $t>> m^{-1}$.

Let us consider at first the early epoch $t<< m^{-1}$, where the gravitational field is strong. For this case the total number density (\ref{n}) of created particles is
\be
n(t)={m^3\over 24\pi^2}\,.
\ee
Notice that in this first approximation, $n$ does not depend on the scale factor $a(t)$ nor on the index $q$. Thus, the particle creation proceeds at just such a rate to keep the particle density constant.

Now let us consider the epoch $t>>m^{-1}$. The total number density of created particles can be split in two components \cite{staro},
\be
n(t)=n_1(t)+n_2(t)\,,\label{nexp}
\ee
where the dominant contribution $n_1$ and the next term $n_2$ in the expansion are given by
\be
n_1(t)=B_{q} m^3 (mt)^{-3q}\,, \hspace{1cm} n_2(t)=C_1 m q^2t^{-2}\,,
\ee
where $B_{q}$ are numerical factors obtained by numeric integration dependent on $q$ and $C_1 = \frac{1}{512\pi}$. According to \cite{staro,gribpav}, $B_{1/2}= 5.3\times 10^{-4}$ and $B_{2/3}= 4.8\times 10^{-4}$ for radiation and matter backgrounds, respectively.

For the total energy density we have, similarly,
\be 
\rho(t)=\rho_1(t)+\rho_2(t)+\rho_3(t)\,,\label{rhoexp}
\ee
with
\begin{eqnarray}
\label{rhos}
\rho_1(t)&=&2mn_1=2B_{q} m^4 (mt)^{-3q}\nonumber\\
\rho_2(t)&=&2mn_2=2C_1 m^2 q^2t^{-2}\\
\rho_3(t)&=&2C_2 m^2 q^2t^{-2}\nonumber
\end{eqnarray}
and $C_2=\frac{1}{48\pi^2}$.

Now, if we consider that the real scalar particles created are of dark matter type, using (\ref{rhodm}) we can find the dark matter creation rate $\Gamma$  for different stages of the evolution, which characterizes a CCDM model.

\section{Constraining dark matter mass}

In order to constrain dark matter particle mass for the present time (matter dominated universe), and take into account the possibility of a back reaction effect in the evolution law close to $t^{2/3}$, we study the behaviour of the above equations for a small deviation of the power $2/3$, namely we make the {\it ansatz} $q=\frac{2}{3}+\varepsilon$. Using Eq. (\ref{rhos}), the Eq. (\ref{rhoexp}) can be rewritten:
\begin{equation}
\rho(t)=2B_qm^4(mt)^{-3q}+2(C_1+C_2)m^2q^2t^{-2}
\label{rhot}
\end{equation}

From the background model (without particle creation), we find the zeroth order expansion rate as $\bar{H}=\frac{\dot{a}}{a}=\frac{q}{t}$, so, we may write
\begin{equation}
\rho(t)=2B_qm^4(mt)^{-3q}+2Cm^2\bar{H}(t)^2=\rho_{cons}+\rho_{pert}
\label{rhoHt}
\end{equation}
where $C=C_1+C_2$. The first term of dark matter density, $\rho_{cons}=2B_qm^4(mt)^{-3q}$, is the conserved part of dark matter, as it corresponds to $\rho_{cons}\propto a^{-3}$. The second term, $\rho_{pert}=2Cm^2\bar{H}(t)^2$, can be seen as the first order correction to the dark matter density, as it does not find correspondence in a model without particle creation.

One could use the Friedmann equation, $\rho=\frac{3H^2}{8\pi G}$, into Eq. (\ref{rhot}) to find $H(t)$. However, this would not yield a full description of quantum particle creation, because the calculated creation rate depends on the background model and {\it vice-versa}, as a back reaction effect, so, we choose to work with an approximation method.

In order to treat the effect of particle creation as a perturbation, we replace the unperturbed $\bar{H}$ by the perturbed $H_1$ and use the Friedmann equation (\ref{fried}) for the perturbed model, $\rho=\frac{3H_1^2}{8\pi G}$ into Eq. (\ref{rhoHt}) to find:
\begin{equation}
H_1(t)=4m\mu\sqrt{\frac{B_q\pi}{3-16\pi C\mu^2}}(mt)^{-\frac{3q}{2}}
\label{eqh1t}
\end{equation}
where $\mu=G^{1/2}m$. This gives an upper limit to the mass: $\mu<\frac{1}{4}\sqrt{\frac{3}{\pi C}}$. Thus, the DM particle must have a mass less than $\approx 4.67$ times the Planck mass ($\approx5.70\times10^{19}$ GeV). 


As one can see from Eq. (\ref{eqh1t}), $H_1\propto t^{-\frac{3q}{2}}$, thus it is equivalent to the model
\begin{equation}
H_1(t)=H_0\left(\frac{t}{t_0}\right)^{-\frac{3q}{2}}
\label{eqh1tt}
\end{equation}
where $H_0$ is the Hubble constant, $t_0$ is the Universe total age. By comparing Eqs. (\ref{eqh1t}) and (\ref{eqh1tt}) we have the following relation
\begin{equation}
\frac{H_0}{m}=\left(\frac{16\pi B_q\mu^2}{3-16\pi C\mu^2}\right)^{-\frac{1}{3\varepsilon}}T^{\frac{2}{3\epsilon}+1}
\end{equation}
where $T=H_0t_0$ is the total age in units of $H_0^{-1}$. This equation relates $m$ (or $\mu$) and $\varepsilon$ to $H_0$ and $T$. Thus, we have three free parameters. Once found 3 parameters from ($\mu$, $\varepsilon$, $H_0$, $T$), the fourth parameter is determined by this equation.

From (\ref{eqh1tt}) we may find the acceleration as:
\begin{equation}
\frac{\ddot{a}}{aH_0^2}=\tau^{-3q}\left(1-\frac{3q}{2T}\tau^{\frac{3\varepsilon}{2}}\right)
\end{equation}
where $\tau=\frac{t}{t_0}$ is the ratio of time to total age. From this equation, one may see that a necessary condition for the existence of a deceleration phase is $q>0$ ($\varepsilon>-\frac{2}{3}$). Furthermore, if $\varepsilon>0$ one has early acceleration, and if $\varepsilon<0$, one has early deceleration, as needed. Thus, $\varepsilon<0$ and $q>0$ are conditions for this model, in such a way that we must find $-\frac{2}{3}<\varepsilon<0$.

From (\ref{eqh1tt}) one may find the scale factor to be
\begin{equation}
a_1(t)=\mathrm{exp}\left[\frac{2T}{3\varepsilon}(1-\tau^{-\frac{3\varepsilon}{2}})\right]
\label{a1t}
\end{equation}
From this equation and using (\ref{fried}) and (\ref{rhodm}), the creation rate can be written as
\begin{equation}
\Gamma(t)=-{(2+3\varepsilon)\over t}+3T\Bigg({t\over t_0}\Bigg)^{-3\varepsilon/2}{1\over t}.
\label{Gamma}
\end{equation}

\begin{figure}[ht!]
    \centering
    \epsfig{file=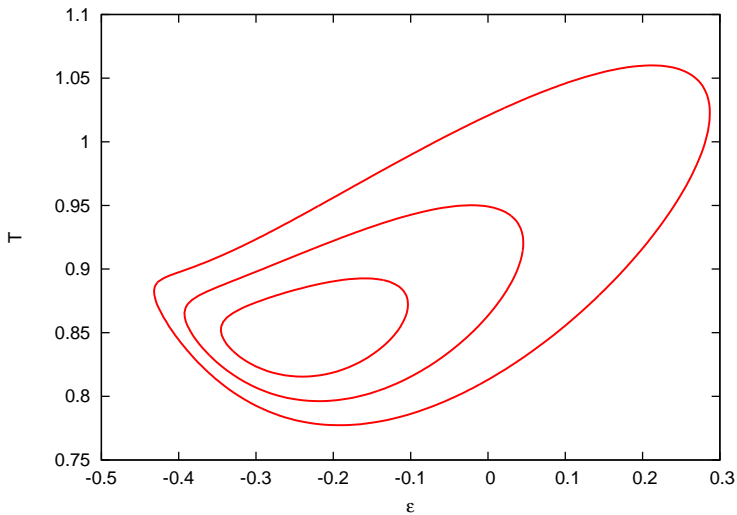,width=10cm}
    \caption{Limits on parameters $\varepsilon$ and $T$ from $H(z)$ data.}
    \label{contoursHt}
\end{figure}

From Eq. (\ref{a1t}), one can find $t(z)$ and replace in (\ref{eqh1tt}) to find $H_1(z)$:
\begin{equation}
H_1(z)=H_0\left[1+\frac{3\varepsilon}{2T}\mathrm{ln}(1+z)\right]^\frac{2+3\varepsilon}{3\varepsilon}
\end{equation}
This can be used to constrain the free parameters through observational $H(z)$ data. We use 28 $H(z)$ compiled data from Farooq and Ratra \cite{FarooqRatra}. By using a $\chi^2$ statistics, where we analytically projected over $H_0$, we remained with only two free parameters ($\varepsilon$ and $T$). The result of the joint analysis can be seen on Fig. \ref{contoursHt}.

As we can see on Fig. \ref{contoursHt}, the $H(z)$ data constrain enough the $\varepsilon-T$ parameter space. We have found the minimum $\chi^2_{\mathrm{min}}=18.00$ for 26 degrees of freedom. The best fit values are $\varepsilon=-0.250^{+0.15}_{-0.096}$$^{+0.30+0.54}_{-0.14-0.18}$ and $T=0.847^{+0.046+0.10}_{-0.031-0.051}$$^{+0.21}_{-0.070}$ at 68.3\%, 95.4\% and 99.7\% c.l. of the joint analysis, respectively.

It should be noticed that this reasonably high value for the best fit $|\varepsilon|$ could imply in a problem for our choice of a perturbative method. However, as one can not provide a full description of the effect of quantum particle creation over the Universe expansion,
we have no means of evaluating this impact. A better situation is for the superior limit on $\varepsilon$ ($-0.10$ at 68\% c.l.) which certainly is in more agreement with a perturbative treatment. This reasonably high value of $|\varepsilon|$ also shows that in order to explain acceleration, a good amount of particle creation is needed.

On Fig. (\ref{Hz}a), we show the $H(z)$ data along with the best fit CCDM model. As one can see, the CCDM model fits well the data in agreement with the low estimated $\chi^2_{\mathrm{min}}$.

\begin{figure}[ht!]
    \centering
    \epsfig{file=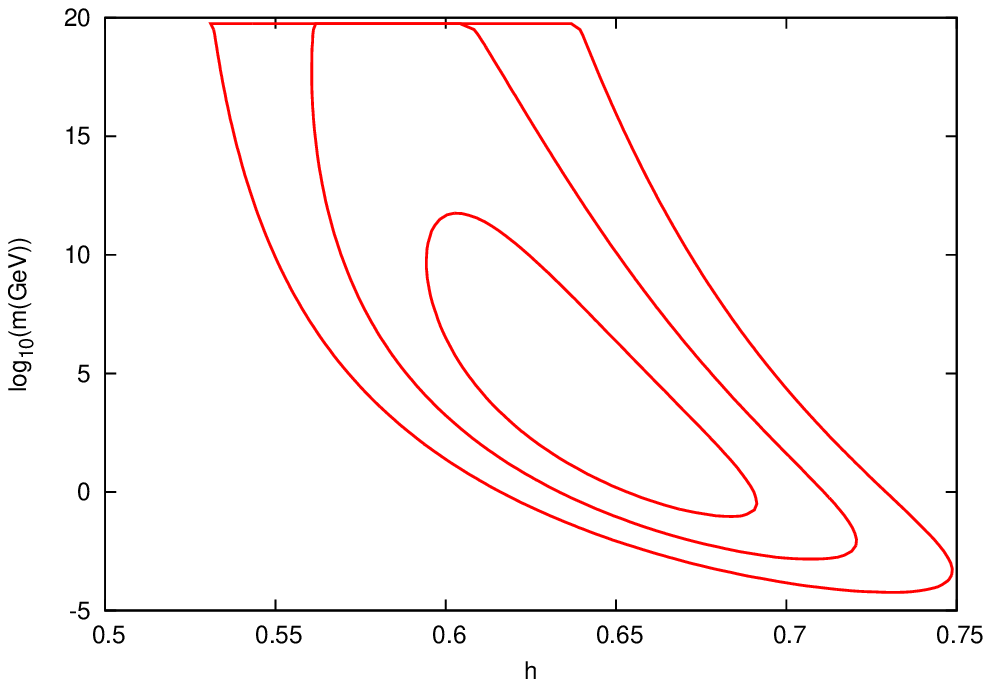,width=8cm}
    \epsfig{file=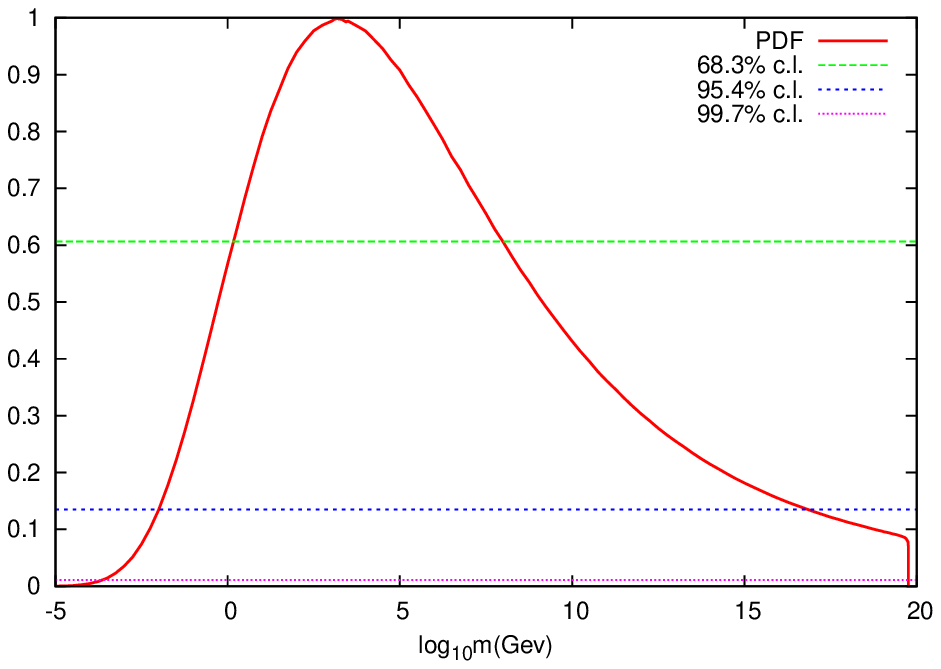,width=8cm}
    \caption{{\bf (a)} Limits on parameters $\log_{10}(m)$ and $h$ from $H(z)$ data, projected over $\varepsilon$. {\bf (b)} Limits on $\log_{10}(m)$ from $H(z)$ data, projected over $\varepsilon$ and $h$.}
    \label{contours}
\end{figure}

However, while examining the parameter space $m$ vs. $\varepsilon$, we have found a strong degeneracy between these parameters. Then, we have examined the plane $m-h$, with projection over $\varepsilon$, where $h\equiv\frac{H_0}{100}(\mathrm{km/s/Mpc})$. As the limit on the mass was much loose, we choose to work with the logarithm of mass, $\log_{10}(m(GeV))$. The result of this joint analysis can be seen on Fig. \ref{contours}a. As on can see on this figure, there is a sudden cut-off for the mass, which correspond to the limit given above, of $5.70\times10^{19}$ GeV. The limit for $h$ was $h=0.644^{+0.031+0.061+0.091}_{-0.033-0.067-0.10}$, which is in agreement with the current limit of $h$ by the Planck, $h=0.673\pm0.012$, in the context of the flat $\Lambda$CDM model \cite{Planck}.

To analyse more closely the limit on the mass, we have projected also over $h$, and have generated the PDF of $\log_{10}(m)$, which can be seen on Fig. \ref{contours}b. As we can see on this figure, there is the same cut-off at $m=5.7\times10^{19}$, but it is beyond the 95.4\% c.l. We have found a best fit for the mass of $1.6\times10^3$ GeV, with ($2.4\times10^{-4}<m<5.7\times10^{19}$) GeV at 99.7\% c.l., ($1.0\times10^{-2}<m<6.7\times10^{16}$) GeV at 95.4\% c.l., and ($1.5<m<6.3\times10^7$) GeV at 68.3\% c.l.

While it seems to be a much loose constraint on DM mass, one should consider that we have used only one experimental data set, namely $H(z)$ data, and we are restricted to the approximated effect of particle creation over the Universe expansion, as explained above.

In order to improve the treatment, reducing the limitations of our model, we could try to use the complete DM density expression (\ref{rhot}) and find a complete Hubble parameter expression taking into account the effects of quantum particle creation. However, it would lead to parametric equations $H(t)$ and $z(t)$, hindering the constraints of $H(z)$ data we intended to do. Other possibility for further improvement would be now consider the corrected $a_1(t)$ (\ref{a1t}) to find a second order effect of particle creation. However, the calculations of particle creation with the background (\ref{a1t}) are much involved, and the analysis turns to be prohibitive.

\section{Comparison with CCDM and $\Lambda$CDM models}
The proposal of phenomenological CCDM models always starts from some specified creation rate function, in general, $\Gamma(H)$. The first proposed CCDM model had $\Gamma\propto H$ \cite{LimaEtAl95,LimaAlcaniz99,AlcanizLima99}, but it was in contradiction with SN Ia observations, as it did not allow for a transition from a decelerated to an accelerated phase. Next, it has been proposed $\Gamma=3\gamma H_0+3\beta H$ \cite{lima01}, solving the problem of no transition. 

Also interesting was the model proposed next, the so called LJO model, with $\Gamma=3\alpha\frac{\rho_{c0}}{\rho_{dm}}H$ \cite{limafernando}, which had the peculiar property of being identical to the concordance flat $\Lambda$CDM model, using a simple equivalence between its parameters. This gave rise to the so called ``dark degeneracy'', where one was unable to distinguish at background level from CCDM to $\Lambda$CDM \cite{limafernando} or even at the linear level of cosmological perturbations \cite{RamosEtAl14}. Put in another words, we could not conclude if the expansion acceleration was due to the quantum vacuum contribution or to the particle creation of DM particles. However, our treatment, relating the particle creation mechanism with the dark matter mass, can be used in the future to solve this ``dark degeneracy''. Better observational data can be used to put strong constraints on dark matter mass within this treatment, which can be compared with other estimates, not depending of the cosmological dynamics, thus concluding if there is such particle creation or not.

Recently, it has been shown the equivalence between CCDM models and $\Lambda(t)$ models \cite{GraefEtAl14}, at the level of background equations. Graef {\it et al.} have analysed three CCDM models, namely, $\Gamma\propto H^{-1}$ (CCDM1), $\Gamma=\gamma$ (constant, CCDM2) and $\Gamma=\frac{c}{H}+\beta H$ (CCDM3). They have shown, in the case of spatial flatness, that the model CCDM1 is equivalent to the LJO model, thereby equivalent to flat $\Lambda$CDM (also shown in \cite{RamosEtAl14}). We will rely on the Graef {\it et al.} analysis in order to make a comparison among our model, CCDM models and $\Lambda$CDM.

In order to compare our particle creation model with the other CCDM models, we rewrite the creation rate as a function of expansion rate, using Eqs. (\ref{Gamma}) and (\ref{eqh1tt}):
\begin{equation}
\Gamma=-\frac{3q}{t_0}\left(\frac{H}{H_0}\right)^{\frac{2}{3q}}+3H
\label{GamaH}
\end{equation}
As one can see, this model has no creation if $q=\frac{2}{3}$ because, in this case, we recover the Einstein-de Sitter model, as expected, where we have $H_0t_0=\frac{2}{3}$. 
It does not recover CCDM1 (or $\Lambda$CDM) or CCDM2 at any value of parameters, but it recovers CCDM3 if $q=-\frac{2}{3}$ and $\beta=3$. However, it would correspond to $\varepsilon=-\frac{4}{3}$, which is far away from the 99.7\% confidence limit of our statistical analysis. Furthermore, as remarked above, $\varepsilon<-\frac{2}{3}$ does not yield deceleration in the past, as required by SN Ia observations \cite{SN}. So, as it does not resemble analytically other known CCDM models, we have compared the models numerically, from low to intermediate redshifts.

\begin{figure}[ht!]
    \centering
    \epsfig{file=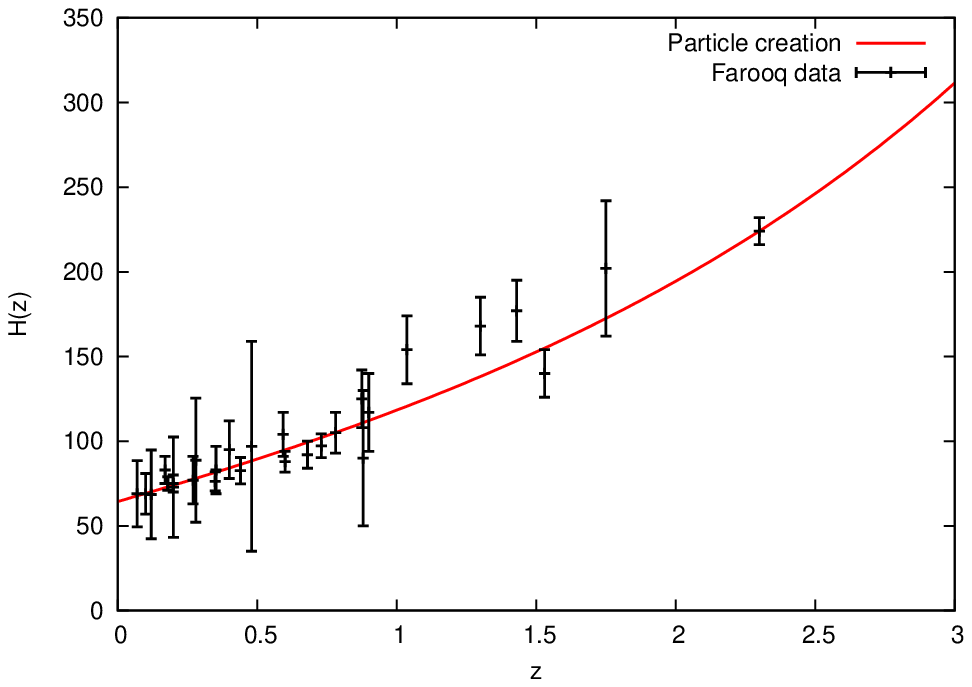,width=8cm}
    \epsfig{file=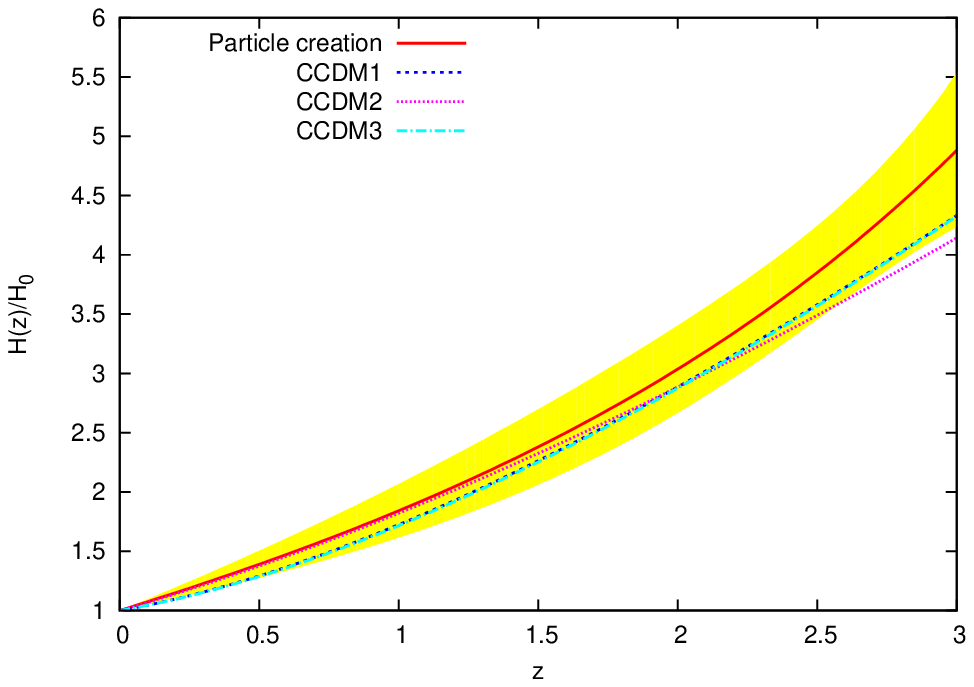,width=8cm}
    \caption{{\bf (a)} Farooq and Ratra \cite{FarooqRatra} $H(z)$ data with best fit particle creation model ($\varepsilon=-0.250$,\, $T=0.847$). {\bf (b)} Comparison of $\frac{H(z)}{H_0}$ among our best fit quantum particle creation model and other 3 best fit CCDM models available in the literature as explained in the text. The shadowed yellow region represents 68\% c.l. deviation from our best fit model.}
    \label{Hz}
\end{figure}

In order to compare numerically the four models, we plot $\frac{H(z)}{H_0}$, on Fig. (\ref{Hz}b), for each model. In our model, we have used the best fit from $H(z)$ data, while for CCDM1, 2 and 3, we used the best fits from \cite{GraefEtAl14}. It is important to notice that the best fits of \cite{GraefEtAl14} come from SNe Ia and CMB/BAO ratio data, not from $H(z)$. We also show the 68\% c.l. deviation region from our model, so, as we can see from Fig. (\ref{Hz}b), all CCDM models are compatible with ours, except for CCDM2. We may also notice that our model tends to give higher expansion rate at high redshifts. However, it can not be seen as a general tendency, as we have used only $H(z)$ data, but better can be seen as an indication of agreement, up to intermediate redshifts, of our model with other already proposed CCDM models.

The great advantage of our treatment really is that it is 
the first attempt of finding a connection between the macroscopic creation rate and the expected quantum particle creation rate, given a reasonable background. With just a few assumptions, we were able to constrain the DM mass. This technique, in the future, with better constraints, can be used to compare with other mass estimates and break the so called ``dark degeneracy''. That is, improvements of this method can be used to decide if it is the quantum vacuum contribution or the quantum particle creation which accelerates the Universe expansion.

\section{Concluding remarks}
In this paper we have analysed a model where a quantum process of particle creation can be responsible for the present acceleration of the universe, the so called CCDM model. Contrary to recent works where the creation rate was only phenomenologically proposed, here we have used a quantum mechanical result for the creation rate of real massive scalar particles in FRW expanding universes in order to constrain the mass of the Dark Matter particle. As far as we know, it is the first time that the results from the quantum mechanical creation process is applied to study the effects on the accelerated expansion of the universe. The quantum creation process admits self corrections to the Hubble expansion rate so that we can interpret this like a kind of back-reaction effect. Such effects could give rise to a small correction $\varepsilon$ of the scale factor on the present matter dominated era. The best fit values found using observational results are $\varepsilon = -0.25$ and $m=1.6\times10^3$ GeV. With such values the model correctly presents a transition from decelerated to accelerated phase, and the particle creation rate is positive near the present epoch. 

It is important to notice that the lower mass limit obtained by this method  ($\sim1 - 10$ GeV), is exactly of the same order of the values expected by experiments of dark matter detectors as DAMA ($\sim15 - 120$ GeV) \cite{dama2008}. The best fit value for the mass can also be compared to different theories of dark matter in the modern cosmological models. According to theories presented in the Introduction, our model correctly predicts the mass as the same order of the mass of LSP, UED and LZP particles, besides satisfying the constraints imposed by the WIMPs model, namely $m>100$ MeV.

Different manners to take into account the self interaction of the field on the expansion rate are possible, leading to different back-reactions effects. Such studies are under investigation. 

\section*{Appendix}

The canonical quantisation of a real scalar field in  curved backgrounds follows in analogy with the quantisation in a flat Minkowski background. The general theory of particle creation from quantum effects in expanding background is developed in details in the books \cite{davies,fulling,mukh2,grib} and also in the articles \cite{partcreation,staro,gribmama02}. Here we will present briefly the general theory for the sake of generality. 

The gravitational metric is treated as a classical external field which is generally non-homogeneous and non-stationary. A real scalar field is electrically neutral and does not couple to the electromagnetic field \cite{mukh2}. This guarantees that such created particles do not interact with radiation, as required for dark matter candidates.

A real scalar field $\phi$ of mass $m$ is described in a curved space-time by the action \cite{davies,fulling,grib,mukh2}
\begin{eqnarray}\label{m63}
S={1\over 2} \int \sqrt{-g}d^4 x \bigg[g^{\alpha\beta}\partial_\alpha\phi(x)\partial_\beta\phi(x)-m^2 \phi^2(x)-\xi R \phi^2(x)\bigg]\,,
\end{eqnarray}
where $\xi$ is a constant parameter that characterises the coupling between the scalar field and the gravitational field and $R$ is the Ricci scalar curvature.  Two values of $\xi$ are of particular interest: the so-called minimally coupled case, $\xi=0$, and the conformally coupled case, $\xi=1/6$. In terms of the conformal time $\eta \equiv \int dt/a(t)$, the metric tensor $g_{\mu\nu}$ is conformally equivalent to the Minkowski metric $\eta_{\mu\nu}$, so that the line element is  $ds^2=a^2(\eta)\eta_{\mu\nu}dx^\mu dx^\nu$, where $a(\eta)$ is the cosmological scale factor. Writing the field $\phi (\eta,\vec{x}) = a(\eta)^{-1}\chi(\eta,\vec{x})$, the equation of motion that follows from the action (\ref{m63}) is
\begin{equation}\label{m67}
\chi''- \nabla^2 \chi +\bigg( m^2a^2+(6\xi -1){a''\over a}\bigg)\chi=0\,,
\end{equation}
where the prime denotes derivatives with respect to the conformal time $\eta$, and we have used $R=6(a''/a^3+\kappa/a^2)$ for the Ricci scalar, with $\kappa$ the spatial curvature for flat ($\kappa=0$), open ($\kappa = -1$) and closed ($\kappa=1$) background.

In which follows we restrict to flat and conformally coupled case, $\kappa=0$ and $\xi=1/6$. The quantisation follows by imposing equal-time commutation relations for the operator $\hat{\chi}$ and its momentum canonically conjugate $\hat{\pi}\equiv\hat{\chi}'$, namely $[\hat{\chi}(\eta,\vec{x}) \,, \hat{\pi}(\eta,\vec{y})]=i\delta(\vec{x}-\vec{y})$. The creation and annihilation operators $\hat{a}_k^\pm$ can be introduced when the field operator $\hat{\chi}$ is expanded as
\be
\hat{\chi}(x,\eta)={1\over 2}\int {d^3k\over (2\pi)^{3/2}}\bigg[\textrm{e}^{ik\cdot x}\chi_k^*(\eta)\hat{a}_k^- + \textrm{e}^{-ik\cdot x}\chi_k(\eta)\hat{a}_k^+\bigg]\,,
\ee
and we find that the mode functions $\chi_k(\eta)$ satisfy a set of decoupled ordinary differential equations \cite{mukh2}
\begin{equation}\label{nv4}
\chi''_k(\eta)+\omega_k^2(\eta)\chi_k(\eta)=0\,,
\end{equation}
with
\begin{equation}\label{m68}
\omega_k^2(\eta)\equiv k^2+m^2a(\eta)^2.
\end{equation}

Each solution $\chi_{k}$ must be normalised for all times according to
\begin{equation}\label{norm}
\chi_{k}(\eta){\chi^*}'_{k}(\eta)-\chi'_{k}(\eta)\chi^*_{k}(\eta)=-2i\,,
\end{equation}
and also they must satisfy the initial conditions
\be
\chi_k(\eta_0)=1/\sqrt{\omega_k(\eta_0)}\,,\hspace{1cm} \chi_k'(\eta_0)=i\sqrt{\omega(\eta_0)}\,.
\ee
Since the point $\eta=0$ is a regular point of Eqs. (\ref{nv4})-(\ref{m68}), we can set $\eta_0=0$.

A standard Bogoliubov transformation \cite{grib,mukh2,staro,gribmama02} introduce the coefficients $\alpha_k$ and $\beta_k$ satisfying $|\alpha_k|^2-|\beta_k|^2=1$ and 
\begin{equation}\label{beta}
|\beta_k|^2={1\over 4\omega_k(\eta)}|\chi'_{k}(\eta)|^2+{\omega_k(\eta)\over 4}|\chi_{k}(\eta)|^2 -{1\over 2}\,.
\end{equation}

The final expression for the total number density of created particles (and antiparticles) is readily obtained by integrating over all the modes,
\be\label{n}
n(\eta)=\bar{n}(\eta)={1\over 2\pi^2a^3}\int_0^\infty k^2 |\beta_k|^2 dk\,,
\ee
where $\bar{n}$ stands for the antiparticles. Similarly, the expression for the total energy density is given by
\be\label{rho}
\rho(\eta)={1\over \pi^2a^4}\int_0^\infty k^2 \omega_k |\beta_k|^2 dk\,,
\ee
By using the inverse transformation of the conformal time $\eta$ it is possible to write the above equations as a function of the physical time $t$.


\begin{acknowledgments}
JFJ is grateful to Unesp - C\^{a}mpus de Guaratinguet\'a for their hospitality while discussing the present work. SHP is grateful to CNPq - Conselho Nacional de Desenvolvimento Cient\'ifico e Tecnol\'ogico, Brazilian research agency, for the financial support, process number 477872/2010-7.
\end{acknowledgments}

\end{document}